# B-DRRN: A Block Information Constrained Deep Recursive Residual Network for Video Compression Artifacts Reduction

Trinh Man Hoang[1]    Jinjia Zhou[1,2]

[1]Graduate School of Science and Engineering, Hosei University, Tokyo, Japan

[2]JST, PRESTO, Tokyo, Japan

*Abstract*—Although the video compression ratio nowadays becomes higher, the video coders such as H.264/AVC, H.265/HEVC, H.266/VVC always suffer from the video artifacts. In this paper, we design a neural network to enhance the quality of the compressed frame by leveraging the block information, called B-DRRN (Deep Recursive Residual Network with Block information). Firstly, an extra network branch is designed for leveraging the block information of the coding unit (CU). Moreover, to avoid a great increase in the network size, Recursive Residual structure and sharing weight techniques are applied. We also conduct a new large-scale dataset with 209,152 training samples. Experimental results show that the proposed B-DRRN can reduce 6.16% BD-rate compared to HEVC standard. After efficiently adding an extra network branch, this work can improve the performance of the main network without increasing any parameters.

## I. INTRODUCTION

Recently, video coders such as H.264/AVC, H.265/HEVC, and the new coming H.266/VVC achieve higher and higher compression ratio. However, all of the lossy compression video encoders involve the compression artifacts, such as blurring artifacts, blocking artifact, color breeding, etc.. Hence, it is essential to get more attention on enhancing the visual quality of the decompressed videos.

Over the past few years, deep learning has made impressive achievements in computer vision and image processing tasks [1] [2] [3]. Especially, C. Dong [4] has introduced SR-CNN to perform the supper resolution task, which is lately useful in enhancing decoded images. Besides that, the authors of the DnCNN [5] demonstrated the multi-tasking ability of Convolutional Neural Network (CNN), which means that the super-resolution (SR) network, the denoising network, and the deblocking network are structurally consistent. Therefore, most super-resolution networks can be used as restoration networks.

To reduce blocking artifacts, Shiba Kuanar [6] introduced a convolutional neural network (CNN) named MDCNN to replace the sample adaptive offset (SAO) of the in-loop filtering part. While based on SR-CNN, C. Dong [7] introduced ARCNN and Y. Dai [8] introduced other VRCNN to replace the whole in-loop filtering part of HEVC. Another RHCNN from T. Wang [9] was applied after both Deblocking and SAO as an additional component of the in-loop filtering. However, since these designs are incompatible with the standards, both the encoder and decoder need to be reimplemented.

Most of the later works prefer process quality enhancement at the decoder end or post-processing, as shown in Fig. 1. Y. Zhang [10] employed DCAD as a post-processing stage without adjusting anything at the compression side. It enhanced decoded frames by using Deep CNN to learn the residual between the decoded frame and the original frame. Although these methods above could improve the quality of compressed videos, their abilities were limited because they did not consider the source of compressed video artifact.

For all the block-based video coding standard, the artifact comes mainly from the block dividing process. Xiaoyi He[11] introduced Double-input CNN to use that block information and got an impressive result. However, it also introduced more parameters, and their network size became much bigger.

To obtain the benefit from this block information without increasing the size of the network, we propose a B-DRRN (Deep Recursive Residual Network with Block information) with a subjective branch for learning block information and a sharing weight technique for keeping the number of model parameters. Therefore, our network can keep a similar size while handling the additional information.

 In summary, we developed a B-DRRN with an additional branch that used the Recursive Residual structure for limiting the size of the network. This branch then is used to learn the features from the block information which is represented for the source of block-based video coding artifact. Furthermore, our B-DRRN uses the sharing weight technique for all branches to keep the number of parameters remains the same with the main branch. The experimental results show that our model can keep a similar number of parameters at the same time of improving the performance of the main branch. This B-DRRN can be applied as a post-processing stage to enhance the quality of any existing block-based compressed video (see Figure 1).

Moreover, we conduct a new large-scale dataset with 209,152 training samples for training such deep CNN as B-DRRN. Our dataset comes from 600 videos with various resolutions and will be published for further research.

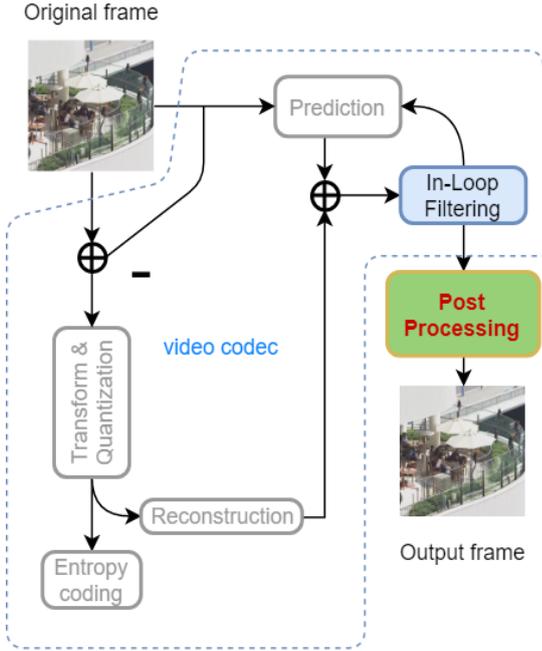

Fig. 1. Post-Processing could be applied to enhance existing compressed video.

## II. THE PROPOSED APPROACH

In this work, we firstly consider the way to represent the block boundary information and design an extra branch to encode this feature; then we designed the B-DRRN for using that representation form. To limit the number of parameters, we use the Recursive Residual structure[12] and sharing weights technique in our design.

### A. Extra branch with block information

The artifact of block-based video coding standard comes from the block dividing process (see Fig 2.a). Because of that, we could use this information to improve the performance of video enhancing process. There are many ways to represent this block information, such as beginning and end position information, boundary information, pixel values vector, etc.… But we need to find one kind of presentation which could distinct the blocks to each other because it is the way that block-based video codec decides to divide the blocks and is also the source of artifacts.

Recently, the Mean mask based method was proven that could get a better result than some other approaches by Xiaoyi He[11] with his Double-inputs CNN (see Fig.2b). This Mean mask frame is conducted by calculating the mean values of all pixels inside a square which starts from the beginning position to the end position of a block. Next, this mean value is assigned back to those pixels inside of that square. The process then repeats over all blocks of the decoded frame. But, to use this method, Double-inputs CNN must add an additional branch, which increases the size of the model a lot. In this work, we use a similar representation for our block information but redesign the additional branch to avoid the increase of the network's parameters.

Our priority when designing this branch is limiting the numbers of increasing parameters as small as possible. We applied the Recursive Residual block from DRRN [12] for this additional branch. This block could repeat many times to help the branch deeper while keeping a small size of the branch;

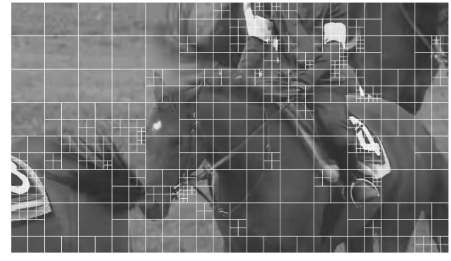

a. Decoded frame with block information

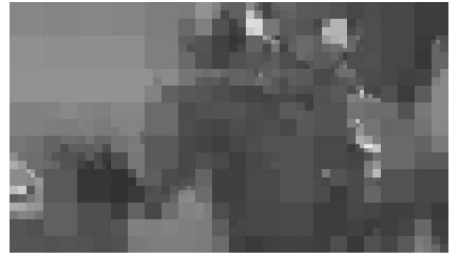

b. Mean mask frame

Fig. 2. Block information representation.

another benefit is it is more stable in training than normal recurrent CNN [13]. Furthermore, to keep intact parameters, we implement sharing weights technique for all the parameters among all branches of our new model. As a result, the parameters remain the same as one branch, and at the same time, the video quality can be enhanced.

It is not trivial for choosing the repeated times for the extra branch; we need to balance its depth with the main branch. In Double-input CNN[11], their additional branch depth is equal $1/4$ of the main branch. However, their two branches used different weight, so the features from their additional branch were not affected by the main branch. In our network, there is a sharing weight process between two branches, so we decided to keep the depth of Subjective branch equal to $1/3$ of the main branch depth. By this design, the extra branch features will be strong enough for not being dominated by the main branch.

### B. Combination of two branches

The output of our two branches then will be fusion. There are two common operators in this situation, which are adding operator and concatenating operator. However, after the fusion process, not only good features arrive but also some bad features still exist. We need to remove the bad features and reorder the remained good features.

Hence, we design a merged branch after the fusion of those two branches features. This merge branch contains two residual layers with the role of the noise canceling and feature reordering unit. Moreover, we also use the sharing weight technique for this branch to keep the number of parameters be the same as only the main branch.

### C. Overall architecture

The input of our network is the decoded frame and its correlative Mean mask frame. Then the output is an enhanced frame with the same size as input. Fig. 3 shows our overall architecture, note that the same color convolution layers share the same weights. Followings are the details of the architecture:

- With the new large-scale dataset, for the more stable training process, a batch normalization layer is first applied for each input and its correlative block mean image.

- The first convolution layer (white color) contains 64 filters with the size of 3x3. And the Recurrent Residual Unit has two convolution layers; each also has 64 3x3 filters.

- At the main branch, the Recurrent Residual unit repeats nine times as DRRN[12] design. As we mentioned in Section II.A, to keep the balance between two branches, the extra branch will have three iterations.

- The outputs of these two branches are combined by adding operator or concatenating operator. For concatenating fusion, the output dimension of the fusion layer will be double, so we need an additional convolution layer goes after to reduce the dimension of the feature from 128 to 64.

- Then, for noise canceling and feature reordering, our merge branch is applied as mention in Section II.B. After that, a 3x3 filter is used for residual reconstruction. Finally, the residual image is added directly to the input for enhancing it.

- As in [14] demonstrated that applying activation function before convolution layer will get better training process. It was also true in the case of DRRN[12]. Hence, in our new model, the Rectifier Linear Unit - ReLu is applied before doing any convolutional calculation.

- After adding the residual to the inputs, when training the network, the Mean Squared Error (MSE) between the original frame and the network's output is used as the loss function.

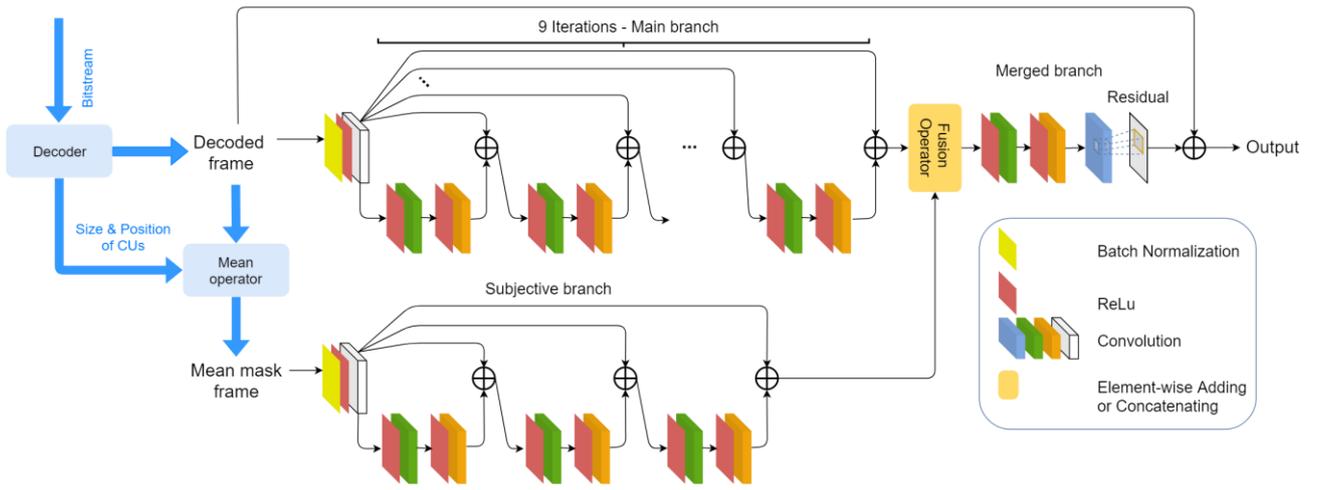

Fig. 3. Our network architecture.

In this architecture, even we add more information into the network, the number of parameters does not change in case of using adding operator. It is satisfied with what we are seeking until now. Another benefit of this approach is that we could add more other information such as motion vector, block type, etc... if we can find a suitable representation for them in the future.

III. EXPERIMENTAL RESULTS

A. Dataset & experimental settings

TABLE I. THE DETAILS OF OUR DATASET

| Resolution | Number of videos |
|---|---|
| Qcif | 140 |
| Cif | 125 |
| 360p | 125 |
| 480p | 105 |
| 720p | 55 |
| Full HD | 40 |
| Ultra HD | 10 |
| **Total** | **600** |

**Dataset.** First, we construct a dataset for learning the network. The training dataset includes 600 sequences with varied video resolutions: qcif, cif, 360p, 480p, HD, full HD, and ultra HD (see table I for more details). All raw video clips are encoded by HM-20.0 at Low-delay P [15] at QPs = 22, 27, 32 and 37. In each pair of the raw clip and decoded clip, we randomly selected four frames to form four training pairs. For each frame pair, we divide them into 64x64 sub-images with a stride of 64, because 64x64 is the biggest block size that HEVC provides, which means the training image could contain the full block structure. We finally get 209,152 sub-image pairs.

**Experimental settings.** Our experiments were conducted on an Ubuntu PC equipped with one Nvidia Tesla V100 GPU. We implement our proposal using PyTorch[16] framework. For training, we use a minibatch size of 256 and Adam[17] optimizer. We start with a learning rate of 5e-04, for stable training, the final layer will have a learning rate equal 0.1*other layers learning rate, then terminate training at 150 epochs. We trained the network at QPs = 22, 27, 32 and 37.

For the evaluation, we test our trained model on five benchmark classes from the common test conditions of HEVC [18], and there is no overlap with the training set. Unlike the training, the test frames are not divided into small patches. Instead, the whole frame is fed into the network. It is because that for convolutional layers, only weights of convolutional filters are learned during training, after training, these filters could be used to convolve the input with any size. The performance of rate-distortion is measured by BD-rate reduction [19] over HM-20.0 with all QPs.

*B. Results and comparison.*

This work is compared with the other four famous models in the field, ARCNN[20], VRCNN[8], DCAD[10], and the baseline model DRRN[12]. To have a fair comparison, all models were trained with the same training set with our approach.

Firstly, we evaluate the performance of rate-distortion by BD-rate reduction over HM-20.0 with four QPs = 22, 27, 32 and 37. Table II compares the BD-rate reduction between all models: ARCNN, VRCNN, DCAD, baseline model DRRN, and our approach, as described in Section II. In table II, the Our_Concat column presents the results of our model under the concatenating operator at the fusion layer, and Our_Adding presents for the adding operator results. From Table II, we can have the following observations:

The first observation is that our approaches (Our_Concat and Our_Adding) achieve better performance than all other methods and when comparing to HM-20.0 software, our approaches can get 6.24% and 6.16% BD-rate reduction reduction more than the baseline DRRN with introducing more parameters (Our_Concat) or 0.93% without introducing any parameters (Our_Adding). This indicates that the information that has been learned from our extra branch can provide more useful information in quality enhancing than DRRN itself. Since the purpose of this work is stressed on the number of parameters, we now just considering Our_Adding, which kept the size of the model to be the same while improving the performance.

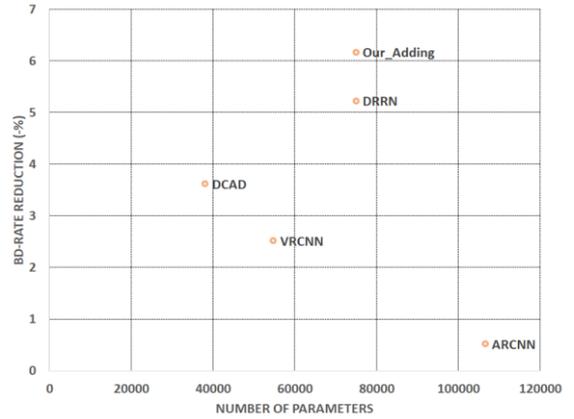

Fig. 4. Our proposal Our_Adding could reduce more BD-rate without adding more parameters.

The second observation is that our approach can get a good effect on any classes of testing videos. Even Our_Adding is not the best compared to Our_Concat; it still has a visible improvement over all classes compared to DRRN and other existing methods. This ability is the result of using the extra branch in learning block information.

Figure 4 shows the relationship between the number of respectively. Especially, it can obtain 1.01% BD-rate parameters and the BD-rate reduction (-%) of all competitive models. Note that the number of parameters here is counted by learnable parameters via PyTorch function and the BD-rate axis is reversed for easy looking. By applying B-DRRN, there is a big gap in BD-rate reduction between Our_Adding and the baseline DRRN but it remains the same on the number of parameters.

TABLE II. COMPARISON OF DIFFERENT METHODS ON BD-RATE REDUCTION (Y-FRAME, %) OVER HM-20.0 BASELINE

| Class | Sequences | ARCNN[7] | VRCNN[8] | DCAD[10] | DRRN[12] | Our_Concat | Our_Adding |
|---|---|---|---|---|---|---|---|
| A | Traffic | -3.03 | -4.14 | -5.69 | -6.85 | **-7.75** | **-7.44** |
| A | PeopleOnStreet | -3.61 | -5.34 | -6.83 | -7.39 | **-7.94** | **-7.95** |
| B | BQTerrace | -1.39 | -3.44 | -3.43 | -5.24 | **-7.17** | **-6.29** |
| B | Kimono | -0.23 | -2.58 | -3.34 | -3.69 | **-4.61** | **-4.88** |
| B | ParkScene | -2.03 | -2.61 | -3.18 | -3.69 | **-4.30** | **-4.34** |
| B | Cactus | -2.61 | -3.37 | -4.96 | -6.65 | **-6.93** | **-7.24** |
| B | BasketballDrive | -0.35 | -0.91 | -3.09 | -4.07 | **-4.94** | **-5.02** |
| C | RaceHourses | -1.93 | -2.51 | -3.98 | -4.26 | **-5.28** | **-5.07** |
| C | BasketballDrill | 1.13 | -1.18 | -4.10 | -6.16 | **-7.46** | **-6.89** |
| C | BQMall | -1.61 | -2.96 | -4.69 | -5.58 | **-6.63** | **-6.47** |
| C | PartyScene | 0.19 | -0.79 | -2.51 | -3.21 | **-4.24** | **-3.92** |
| D | BQSquare | 1.25 | -1.42 | -3.30 | -5.73 | **-6.90** | **-6.59** |
| D | BlowingBubles | 0.67 | -1.00 | -3.15 | -3.85 | **-5.09** | **-4.84** |
| D | BasketballPass | -1.33 | -2.08 | -3.93 | -3.84 | **-5.43** | **-5.14** |
| D | RaceHorses | -2.90 | -3.43 | -4.99 | -5.25 | **-6.40** | **-6.09** |
| E | Johnny | 2.48 | -1.10 | 0.61 | -3.62 | **-4.82** | **-5.47** |
| E | FourPeople | -3.14 | -4.80 | -6.45 | -8.79 | **-9.45** | **-9.62** |
| E | KristenAndSara | -1.30 | -3.19 | -3.90 | -6.36 | **-6.98** | **-7.71** |
| Average | | -1.10 | -2.60 | -3.94 | -5.23 | **-6.24** | **-6.16** |

IV. CONCLUSION

This paper presents a recursive residual learning approach for using additional information in HEVC. It is challenging to keep learning model is not heavier to use additional information but still could enhance the performance. These challenges come from the choice of a suitable structure for model designing, how to keep the balancing between the main features and the additional features, how to represent the additional features, and so on. This model uses the block information, including the boundary information and a mean value of the block. With our idea of weight sharing between two branches and cautious balancing decision, our model could get over all challenges. Experimental results show that our approach can get better videos quality with an average of 6.16% BD-rate compared to HEVC standard. Furthermore, the B-DRRN can be extended to any further additional information in the future.

ACKNOWLEDGMENT

This work is supported by JST, PRESTO Grant Number JPMJPR1757 Japan.